\documentclass[preprintnumbers,floats,twocolumn,prd,aps]{revtex4}
\usepackage{amssymb,amsmath,epsfig}

\newcommand{\K}{{\cal K}}\renewcommand{\vec}[1]{\mbox{\boldmath$#1$}}

\begin{document}

\title{Is It Really Naked?\\On Cosmic Censorship in String Theory}
\author{Andrei V. Frolov}\email{afrolov@stanford.edu}
\affiliation{
  KIPAC/SITP, Stanford University\\
  Stanford, CA, 94305-4060
}
\date{September 10, 2004}

\begin{abstract}
  We investigate the possibility of cosmic censorship violation in string
  theory using a characteristic double-null code, which penetrates
  horizons and is capable of resolving the spacetime all the way to the
  singularity. We perform high-resolution numerical simulations of the
  evolution of negative mass initial scalar field profiles, which were
  argued to provide a counterexample to cosmic censorship conjecture
  for AdS-asymptotic spacetimes in five-dimensional supergravity. In no
  instances formation of naked singularity is seen. Instead, numerical
  evidence indicates that black holes form in the collapse. Our results
  are consistent with earlier numerical studies, and explicitly show
  where the `no black hole' argument breaks.
\end{abstract}

\preprint{SU-ITP-04-17}
\maketitle

\section{Introduction}\label{sec:into}

Cosmic censorship conjecture remains an unsolved problem in general
relativity. In the absence of proof, finding an acceptable
counterexample is very important, as it would resolve the issue one way
or the other. But even that remains elusive. While there are numerous
examples of initial conditions that form a naked singularity in general
relativity, none of them are generic as required by the terms of the
cosmic censorship conjecture.

Recently, Hertog, Horowitz, and Maeda argued that cosmic censorship is
generically violated in asymptotically anti de Sitter spacetimes
\cite{Hertog:2003zs}, in particularly in five-dimensional supergravity
arising from string theory constructs \cite{Hertog:2003xg}. The
counterexample consists of an initial data set which is known to form a
singularity and an argument that it cannot form a black hole in the
collapse. That would leave the possibilities that singularity is naked,
which would disprove cosmic censorship conjecture, or consumes the
entire spacetime (big crunch), which initially was deemed unlikely by
the authors.

These papers have caused quite a controversy in the literature. The
arguments of \cite{Hertog:2003zs, Hertog:2003xg} were called into
question by a number of papers, on grounds of both analytical
\cite{Dafermos:2004ws, Hubeny:2004cn} and numerical
\cite{Gutperle:2004jn, Garfinkle:2004sx, Garfinkle:2004pw}
calculations. The authors themselves pointed out a gap
\cite{Hertog:2004gz} in their first example \cite{Hertog:2003zs}, and
reconsidered the possibility of a big crunch \cite{Hertog:2004rz}.

The present paper investigates the possibility of cosmic censorship
violation numerically. Earlier numerical studies \cite{Gutperle:2004jn,
Garfinkle:2004sx, Garfinkle:2004pw} are suggestive of a black hole
formation. However, they are not conclusive due to the flawed
coordinate choice, which breaks at the moment of apparent horizon
formation, and prevents the code from seeing the whole spacetime,
thus leaving the possibility of a singularity subsequently becoming
naked. This shortcoming is resolved in the present paper by using a
characteristic double-null code.

Conceptually, the code is similar to the one used to investigate the 
global structure of the spacetime in the critical collapse of the
scalar field \cite{Frolov:2003dk}, although the implementation is
entirely new. Several issues specific to asymptotically AdS spacetimes
had to be addressed, in particular boundary conditions at infinity.

We focus on cosmic censorship counterexample in the string theory
context \cite{Hertog:2003xg}, and perform high-resolution numerical
simulations of the evolution of two families of negative mass initial
scalar field profiles for various values of parameters and different
boundary conditions at the cutoff. In no instances formation of naked
singularity is seen. Instead, numerical evidence indicates that black
holes form in the collapse, in agreement with \cite{Gutperle:2004jn}.
Our results clearly show where the `no black hole' argument of
\cite{Hertog:2003xg} fails.

The paper is organized as follows: In Section~\ref{sec:coord}, the
issue of a coordinate choice is discussed in view of the requirements
of testing cosmic censorship conjecture numerically. The scalar field
evolution equations and the implementation of the algorithm are
described in Section~\ref{sec:evolution}. The numerical results of
investigation of cosmic censorship in string theory are given in
Section~\ref{sec:string}, and summarized in Section~\ref{sec:conlusion}.

\section{Coordinate Choice}\label{sec:coord}

In numerical investigations of singularity formation and global
structure of the spacetime, the coordinate choice plays a crucial role.
To study the global structure, coordinates must cover the entire
spacetime. To see the singularity inside a black hole, coordinates must
penetrate the horizon. To prevent the singularity from corrupting the
rest of the spacetime once it forms (if it is not naked, that is), the
information propagation speed (both numerical and true characteristic)
must be under control. These issues cannot be avoided if one wishes to
study cosmic censorship numerically.

In this paper, we adopt double-null coordinate system, which is
particularly suited for the above requirements. The spherically
symmetric $(n+2)$-dimensional spacetime metric can be written as
\begin{equation}\label{eq:conf}
  ds^2 = e^{-2\sigma} d\eta^2 + r^2 d\Omega_n^2,
\end{equation}
where $d\Omega_n^2$ is the metric of a unit $n$-dimensional sphere, and
the two-manifold metric $d\gamma^2 = e^{-2\sigma} d\eta^2$ is written
in explicitly conformally-flat form. Throughout the paper, we will be
freely switching between Minkowski and double-null coordinates on the
flat two-manifold
\begin{equation}\label{eq:flat}
  d\eta^2 = -dt^2 + dx^2 = -4\, du\,dv,
\end{equation}
with $u = (t-x)/2$ and $v = (t+x)/2$.
The familiar form of anti de~Sitter metric in spherically symmetric
static coordinates
\begin{equation}\label{eq:ads}
  ds^2 = - \left({\textstyle 1+\frac{r^2}{\ell^2}}\right) dt^2 + \frac{dr^2}{1+\frac{r^2}{\ell^2}} + r^2 d\Omega_n^2
\end{equation}
transforms to
\begin{equation}
  r = \ell \tan\frac{x}{\ell}, \hspace{1em}
  \sigma = \ln\left(\cos\frac{x}{\ell}\right)
\end{equation}
in coordinate system (\ref{eq:conf}). One should note that the peculiar
property of AdS that the spatial infinity ``is not very far away'' is
reflected in the fact that a finite interval $x \in [0,\frac{\pi}{2}\,\ell]$
covers the whole AdS spacetime in conformal coordinates (\ref{eq:conf}).
Indeed, these are the coordinates used in construction of a
Carter-Penrose conformal diagram of the AdS spacetime.

In spherical symmetry, one can define a local mass by
\begin{equation}\label{eq:m}
  f \equiv g^{\mu\nu}r_{,\mu}r_{,\nu} = 1 - \frac{2m}{r^{n-1}}.
\end{equation}
The function $f$ carries information about the spacetime structure: it
is negative in the trapped (or anti-trapped) region, positive in the
regular region, and vanishes on the apparent horizon. The mass function
defined by (\ref{eq:m}) coincides with ADM mass in asymptotically flat
spacetimes. In asymptotically AdS spacetimes, however, it diverges near
spatial infinity. The physical cause for this is simple: AdS has a
constant (negative) energy density while the volume is infinite. To
avoid infinities, one can subtract the divergent part from the
definition of the mass
\begin{equation}\label{eq:mu}
  \mu = m + \frac{1}{2} \frac{r^{n+1}}{\ell^2},
\end{equation}
where $\ell$ is the curvature radius of the asymptotic AdS.
We will refer to this mass definition as the {\em reduced mass}.

Although traditionally used in numerical relativity, Schwarzschild
coordinates, like the ones in metric (\ref{eq:ads}), are not suited to
study the possible violation of cosmic censorship in dynamical
evolution. They do not penetrate horizons (at the apparent horizon,
$f=g^{rr}=0$, so the metric coefficient $g_{rr}$ necessarily diverges),
and therefore have no chance of seeing, for example, the destruction of
a black hole by infall of a negative mass and ``baring'' of the
singularity inside (if it actually happens). Unfortunately,
Schwarzschild coordinates were the coordinates used in recent numerical
studies \cite{Gutperle:2004jn, Garfinkle:2004sx, Garfinkle:2004pw} of
the possible counterexamples to cosmic censorship in asymptotically
AdS spacetimes \cite{Hertog:2003zs, Hertog:2003xg}. Although the
formation of a trapped surface is seen, the further fate of the
spacetime remains unknown. This renders the results of
\cite{Gutperle:2004jn, Garfinkle:2004sx, Garfinkle:2004pw} largely
inconclusive as far as the possible violation of the cosmic censorship
goes.

There are coordinate systems other than (\ref{eq:conf}) which do
penetrate the horizon, most notably the ingoing Eddington-Finkelstein
coordinates
\begin{equation}\label{eq:efv}
  ds^2 = -f e^{2g} dv^2 + 2e^g dv dr + r^2 d\Omega_n^2.
\end{equation}
However, in implementation of the evolution code based on this
coordinate system (which we did as well, although it is not going to be
further discussed in this paper) one encounters a problem that the
numerical information propagation speed is {\em higher} than the true
speed of the outgoing characteristic (which varies on the grid). This
leads to termination of the numerical evolution in the regions causally
disconnected from the singularity, which is not a satisfactory behavior
for a code that aspires to resolve the global structure of the spacetime.

The true strength of double-null coordinate system (\ref{eq:conf}) is
the advantage of knowing the speed of information propagation {\em
everywhere in advance}. The importance of this point cannot be
overemphasized, as numerical information propagation properties is what
makes or breaks the numerical code which has to deal with singularities
before any other factor even comes into play.

\begin{figure}
  \centerline{\epsfig{file=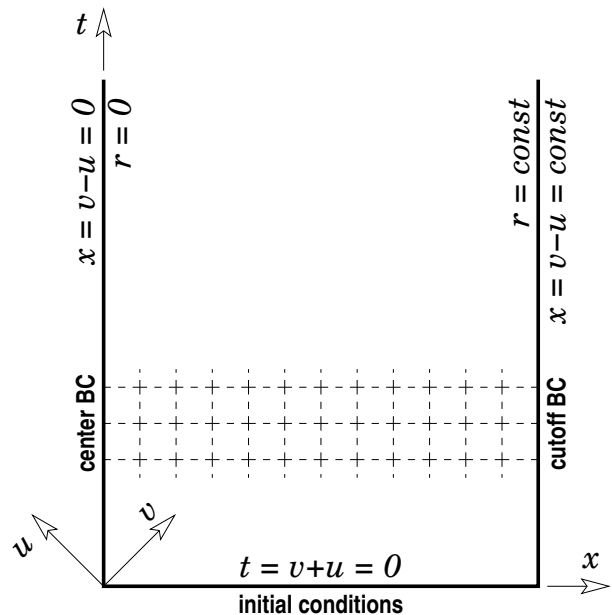, width=8cm}}
  \caption{Coordinate choice.}
  \label{fig:gauge}
\end{figure}

The final issue that needs to be discussed before we go over to
evolution equations is the gauge choice. The form of the metric
(\ref{eq:conf}) is left invariant under rescaling of null coordinates
\begin{equation}\label{eq:gauge}
  u \mapsto U(u), \hspace{1em}
  v \mapsto V(v).
\end{equation}
We use this gauge freedom to place initial and boundary surfaces at
known coordinate locations, namely fix center of spacetime and
constant-$r$ cut-off at $x=0$ and $x=c$ correspondingly, and put
initial data surface at $t=0$ timeslice, as illustrated in Fig.~\ref{fig:gauge}.
This can always be achieved simply by putting $u=v$ at $r=0$ and $u=-v$
at initial surface, while rescaling $v$ to have $v-u=c$ at cut-off. In
fact, it does not even fix the gauge uniquely. The residual gauge
freedom (\ref{eq:gauge}) is determined by
\begin{subequations}\label{eq:freedom}
\begin{eqnarray}
  V(u) - U(u) = 0, && u > 0,\\
  V(u+c) - U(u) = c, && u > -c,\\
  V(-u) + U(u) = 0, && 0 \ge u \ge -c,
\end{eqnarray}
\end{subequations}
and amounts to an odd periodic function $\omega$ of period $c$, with
$V(u)=U(u)=u+\omega(u)$.

\section{Scalar Field Evolution}\label{sec:evolution}

The dynamics of a scalar field $\phi$ with potential $V(\phi)$
minimally coupled to gravity in $N=2+n$ dimensions is described
by the action
\begin{equation}\label{eq:action}
  S = \frac{1}{16\pi G} \int 
           \big\{R - g^{\mu\nu}\phi_{,\mu}\phi_{,\nu} - 2V(\phi)\big\}\, \sqrt{-g}\, d^N x.
\end{equation}
Spherical symmetry effectively reduces the problem to evolution of
three non-linearly coupled scalar fields on a flat two-dimensional
manifold. Substituting the spherically symmetric metric (\ref{eq:conf})
into the scalar field action (\ref{eq:action}) and integrating over the
$n$-dimensional spherical subspace, one obtains the reduced action
\begin{eqnarray}\label{eq:reduced}
  S &\propto& \int \Big\{
      {\textstyle \frac{n(n-1)}{r^2}} \left[(\nabla r)^2 + e^{-2\sigma}\right]
      - {\textstyle \frac{2n}{r}} (\nabla r \cdot \nabla\sigma) \nonumber\\
    && \hspace{2.5em}
      - (\nabla\phi)^2 - 2 e^{-2\sigma} V(\phi)
    \Big\}\, r^n\, d^2 x,
\end{eqnarray}
which describes the dynamics of a spherically symmetric gravitating
scalar field. Here and later the differential operators (gradient
$\nabla$ and D'Alembertian $\Box$) are taken with respect to the flat
two-dimensional metric (\ref{eq:flat}). Variation of the reduced action
with respect to field variables $\sigma$, $\phi$, and $r$ gives
equations of motion
\begin{subequations}\label{eq:eom}
\begin{eqnarray}
  \frac{r\Box r}{n-1} + (\nabla r)^2 &=&
    e^{-2\sigma} \left(1 - {\textstyle \frac{2r^2 V}{n(n-1)}}\right), \label{eq:eom:r}\\
  \Box\phi + \frac{n}{r}\, (\nabla r \cdot \nabla\phi) &=& e^{-2\sigma} V', \label{eq:eom:phi}\\
  \Box\sigma - \frac{n}{2} \frac{\Box r}{r} - \frac{1}{2} (\nabla\phi)^2 &=&
    \frac{2}{n}\, e^{-2\sigma} V, \label{eq:eom:sigma}
\end{eqnarray}
\end{subequations}
while constraint equations
\begin{equation}\label{eq:constraint}
  \text{Traceless}\left[r_{,ab} + 2\,r_{(,a}\sigma_{,b)} + {\textstyle \frac{r}{n}}\, \phi_{,a}\phi_{,b}\right] = 0
\end{equation}
are recovered by variation with respect to the (flat) two-metric
$\eta_{ab}$. The traceless part of a symmetric two-tensor has two
independent components, which means there are two constraints. They
can be taken either as $\{uu\}$ and $\{vv\}$ or $\{xt\}$ and
$\{xx\}+\{tt\}$ components of (\ref{eq:constraint}).

The evolution code uses a semi-constrained algorithm. Rather than solve
dynamical equation (\ref{eq:eom:sigma}) directly, ingoing ($uu$) null
constraint equation
\begin{equation}\label{eq:constraint:uu}
  r_{,uu} + 2\,r_{,u}\sigma_{,u} + {\textstyle \frac{r}{n}}\, \phi_{,u}^2 = 0
\end{equation}
is integrated to obtain one of the metric coefficients. The integration
is particularly easy to do if one defines a new field variable
\begin{equation}\label{eq:g}
  g = -2\sigma - \ln(-r_{,u}),
\end{equation}
in terms of which the constraint (\ref{eq:constraint:uu}) is written simply as
\begin{equation}\label{eq:constraint:g}
  g_{,u} = \frac{r}{n} \frac{\phi_{,u}^2}{r_{,u}}.
\end{equation}
The field variable $g$ actually has a meaning beyond a mere
computational convenience. It is the same $g$ which enters coefficients
of the Eddington-Finkelstein metric (\ref{eq:efv}).

Equation (\ref{eq:eom:r}) warrants further discussion. It involves a
delicate cancellation of terms in both small $r$ and large $r$ regimes,
which makes it susceptible to discretization errors. The loss of
precision can be catastrophic, leading to numerical instability
developing at either $r=0$ or large $r$ and subsequent code failure.
After trying various discretization schemes, we settled on discretizing
a form of equation (\ref{eq:eom:r}) which eliminates non-linear
gradient term in favour of a box operator for small $r$
\begin{equation}
  \frac{\Box(r^n)}{n(n-1)} = -r^{n-2} r_{,u} e^g \left(1 - {\textstyle \frac{2 r^2 V}{n(n-1)}}\right),
\end{equation}
while for large $r$, the equation (\ref{eq:eom:r}) re-written in terms
of inverse radius $\rho=1/r$ is discretized instead
\begin{equation}
  \rho\Box\rho - (n+1)(\nabla\rho)^2 = -\rho_{,u} e^g \left[(n-1)\rho^2 - {\textstyle \frac{2}{n}} V\right].
\end{equation}
This largely circumvents the stability problems mentioned, while still
being possible to discretize efficiently.

The Cauchy problem for evolution equations (\ref{eq:eom}) requires
specification of six functions of one variable at the initial spacelike
surface: the values of three fields $r,\sigma,\phi$ and their time
derivatives $\dot{r},\dot{\sigma},\dot{\phi}$. These six functions are
not independent in general relativity; they must satisfy two constraint
equations (\ref{eq:constraint}). We restrict our attention to
time-symmetric initial data. In covariant form, the requirement of time
symmetry is written as
\begin{equation}
  \K_{ab} = 0, \hspace{1em}
  \vec{n} \cdot \nabla\phi = 0,
\end{equation}
where $\K_{ab}$ is an extrinsic curvature and $\vec{n}$ is the normal
to the initial data surface. In our gauge, the initial surface is
$t=0$, so the condition of time symmetry is simply that the time
derivatives $\dot{r},\dot{\sigma},\dot{\phi}$ vanish. This requirement
satisfies the $\{tx\}$ constraint identically, which leaves three
functions subject to one constraint for specification of the initial
data. Of the two freely specifiable functions, one is physical - the
initial scalar field profile, and the other is a gauge choice fixing
residual gauge freedom (\ref{eq:freedom}).

Once the scalar field profile is specified on the initial time slice,
the rest of the variables are obtained by integrating
\begin{subequations}\label{eq:ic}
\begin{eqnarray}
  r_{,x} &=& \left(1 - \frac{2m}{r^{n-1}}\right) e^g, \label{eq:ic:r}\\
  m_{,r} + m r \frac{\phi_{,r}^2}{n} &=& \frac{r^n}{n} \left(V + \frac{\phi_{,r}^2}{2}\right), \label{eq:ic:m}\\
  g_{,r} &=& \frac{r}{n}\, \phi_{,r}^2 \label{eq:ic:g}.
\end{eqnarray}
\end{subequations}
Equation (\ref{eq:ic:r}) follows from definitions (\ref{eq:m},\ref{eq:g}).
Equation (\ref{eq:ic:m}) is the remaining constraint equation
re-written in terms of the physically interesting mass function $m$.
Our gauge choice is implicitly given by equation (\ref{eq:ic:g}); it
corresponds to the requirement $\dot{g}=0$ on the initial time slice.
The integration is performed numerically using fourth order Runge-Kutta
algorithm with constant step size.

Since in asymptotically AdS spacetimes the light rays take finite time
to reach spatial infinity and reflect back, the issue of boundary
conditions at infinity is a physical one, and cannot be avoided.
Various boundary conditions for scalar field $\phi$ at infinity have
been suggested, specifics of which we will discuss later. Additional
complication for numerical evolution is that it is unfeasible to
include infinity on the grid, so the spacetime has to be cut-off, which
we do at a constant (large) radius $r_c$, and impose boundary conditions
for the scalar field there. The remaining boundary condition at the
cut-off is the one for the metric coefficient
\begin{equation}\label{eq:bc:8}
  g_{,t} = \frac{\rho_{,xt}}{\rho_{,x}} - \frac{2}{n} \frac{\rho\, \phi_{,t} \phi_{,x}}{\rho_{,x}},
\end{equation}
which follows from the $\{tx\}$ constraint equation (\ref{eq:constraint}).
In the center of the spacetime, one has the usual regularity condition
for the scalar field
\begin{equation}\label{eq:bc:0}
  \nabla r \cdot \nabla\phi = 0.
\end{equation}

\begin{figure}
  \centerline{\epsfig{file=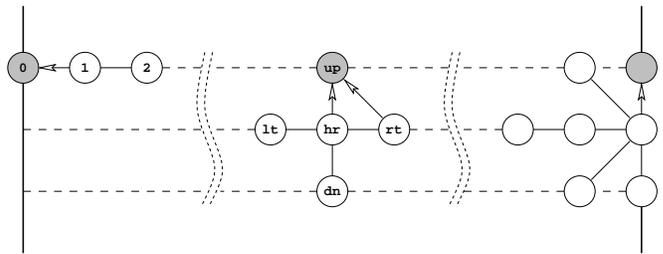, width=\columnwidth}}
  \caption{Numerical evolution scheme.}
  \label{fig:grid}
\end{figure}

The numerical evolution scheme is illustrated in Fig.~\ref{fig:grid}.
The field variables are discretized on a square spacetime grid of
spacing $\epsilon$. Discretization of the evolution equations is done
using the leapfrog scheme, which is second order accurate and
non-dissipative. In particular, the gradient and box differential
operators are discretized as
\begin{subequations}
\begin{equation}
  \nabla_{\!x} X = \frac{X_{\text{rt}} - X_{\text{lt}}}{2\epsilon}, \hspace{1em}
  \nabla_{\!t} X = \frac{X_{\text{up}} - X_{\text{dn}}}{2\epsilon},
\end{equation}
\vspace{-1.5em}
\begin{equation}
  \Box X = \frac{1}{\epsilon^2} (X_{\text{rt}} + X_{\text{lt}} - X_{\text{up}} - X_{\text{dn}}).
\end{equation}
\end{subequations}
Neumann boundary conditions (\ref{eq:bc:0}) require asymmetric
discretization of the derivatives for second order accuracy
\begin{equation}
  X'(0) = \frac{1}{2\epsilon} (4X_1 - 3X_0 - X_2).
\end{equation}
The code was implemented in Fortran, and is efficient enough to run on
personal workstation class computers. Since only three subsequent time
steps are stored, the code is not memory-limited even for very large
grids. All the numerical results presented in this paper were obtained
in simulations with the grid size of 65539 points.

\section{Cosmic Censorship in String Theory}\label{sec:string}

Hertog, Horowitz, and Maeda argued that cosmic censorship is
generically violated in string theory \cite{Hertog:2003xg}. In their
counterexample, they considered ${\cal N}=8$ gauged supergravity in
five dimensions, which is thought to be a consistent truncation of ten
dimensional type IIB supergravity on $S^5$. They picked a single scalar
component which does not act as a source for any of the other fields,
thereby reducing the problem to the one of Einstein gravity in five
dimensions minimally coupled to a scalar field $\phi$ with negative
potential
\begin{equation}\label{eq:5D:V}
  V(\phi) = -2 e^{2\phi/\sqrt{3}} - 4 e^{-\phi/\sqrt{3}}.
\end{equation}
The potential is unbounded from below, with a maximum value of
$V(0)=-6$ which corresponds to the asymptotically AdS spacetime with
curvature scale $\ell=1$. They constructed time-symmetric initial data
which has negative reduced mass $\mu_0$ of the initial configuration,
namely
\begin{equation}\label{eq:5D:IC}
  \phi = \left\{\begin{array}{c@{,\hspace{2em}}c}
    {\displaystyle \frac{A}{R_0^2}} & r\le R_0 \medskip\\
    {\displaystyle \frac{A}{r^2}} & r > R_0 \\
  \end{array}\right. .
\end{equation}
They further argued that a black hole cannot form in the collapse of
this initial configuration (as Schwarzschild-AdS black holes have
positive mass), whereas it is possible to show that singularity does
form in the collapse of the homogeneous part of the field, and hence
concluded that singularity must be naked and cosmic censorship
conjecture is violated.

We use the numerical code described in the previous section to solve
Einstein and scalar field equations for the spherically symmetric
collapse of the profile (\ref{eq:5D:IC}), and determine the global
structure of the resulting spacetime. We find that black holes {\em do}
form in the collapse of initial data (\ref{eq:5D:IC}), contrary to the
argument of Hertog {\it et.~al.} \cite{Hertog:2003xg}.

We place the spacetime cut-off at $r_c=25$, which is large enough so
that less than one percent of the initial profile mass $\mu_0(\infty)$ lies
beyond the cut-off in all cases. At the cut-off, we impose either
Dirichlet or Neumann boundary conditions for the scalar field
\begin{subequations}
\begin{eqnarray}\label{eq:5D:BC}
  \phi\big|_{r=r_c} = \text{const}, && \text{[Dirichlet BC]} \label{eq:5D:BC:D} \\
  \partial_r (r^2\phi)|_{r=r_c} = 0, && \text{[Neumann BC]} \label{eq:5D:BC:N} 
\end{eqnarray}
\end{subequations}
Neumann boundary conditions (\ref{eq:5D:BC:N}) correspond to the
``standard'' boundary conditions on the fall-off of the scalar field
$\phi = \alpha(t)/r^2$ in AdS-CFT correspondence, but are written in
this form for computational convenience. We also try alternative
suggestion \cite{Hertog:2003xg} for the initial field profile,
\begin{equation}\label{eq:5D:IC:ln}
  \phi = \left\{\begin{array}{c@{,\hspace{2em}}c}
    {\displaystyle A\, \frac{\ln R_0}{R_0^2}} & r\le R_0 \medskip\\
    {\displaystyle A\, \frac{\ln r}{r^2}} & r > R_0 \\
  \end{array}\right. ,
\end{equation}
which has a slower fall-off rate and much larger negative masses than
(\ref{eq:5D:IC}).

\begin{table*}
  \begin{center}
    \begin{tabular}{||r|r|r|r||r|r||r|r|r||}
      \hline
      \multicolumn{9}{||c||}{\normalsize truncated $1/r^2$ field profile (\ref{eq:5D:IC})} \\
      \hline\hline
      \multicolumn{4}{||c||}{initial conditions} & 
      \multicolumn{2}{c||}{Dirichlet BC} &
      \multicolumn{3}{c||}{Neumann BC} \\
      \hline
      $\phi_0$ & $A$ & $R_0$ & $\mu_0$ &
      $r_\text{BH}$ & $\mu_\text{BH}$ &
      $r_\text{BH}$ & $\mu_\text{BH}$ & $\mu_\text{cut}$ \\
      \hline
	0.5 &  2.0 & 2.0 &   -0.50 & 0.86 & 0.64 & 0.92 & 0.78 & 0.76 \\
	1.0 &  4.0 & 2.0 &   -2.42 & 1.28 & 2.16 & 1.40 & 2.90 & 2.81 \\
	1.5 &  6.0 & 2.0 &   -5.74 & 1.60 & 4.56 & 1.75 & 6.22 & 5.88 \\
	0.5 &  4.5 & 3.0 &   -3.45 & 1.33 & 2.45 & 1.46 & 3.34 & 3.23 \\
	1.0 &  9.0 & 3.0 &  -15.58 & 1.88 & 8.01 & 2.08 & 11.5 & 11.2 \\
	1.5 & 13.5 & 3.0 &  -35.65 & 2.32 & 17.2 & 2.57 & 25.1 & 23.5 \\
	0.5 &  8.0 & 4.0 &  -11.89 & 1.78 & 6.60 & 1.97 & 9.47 & 9.39 \\
	1.0 & 16.0 & 4.0 &  -52.78 & 2.47 & 21.7 & 2.74 & 31.9 & 31.9 \\
	1.5 & 24.0 & 4.0 & -119.51 & 3.05 & 47.9 & 3.38 & 71.0 & 68.0 \\
      \hline
    \end{tabular}
    \hspace{2cm}
    \begin{tabular}{||r|r|r|r||r|r||}
      \hline
      \multicolumn{6}{||c||}{\normalsize truncated $\ln(r)/r^2$ field profile (\ref{eq:5D:IC:ln})} \\
      \hline\hline
      \multicolumn{4}{||c||}{initial conditions} & 
      \multicolumn{2}{c||}{Dirichlet BC} \\
      \hline
      $\phi_0$ & $A$ & $R_0$ & $\mu_0$ &
      $r_\text{BH}$ & $\mu_\text{BH}$ \\
      \hline
	0.5 &  2.885 & 2.0 &  -25.1  & 0.88 & 0.69 \\
	1.0 &  5.771 & 2.0 & -103.5  & 1.34 & 2.51 \\
	1.5 &  8.656 & 2.0 & -238.1  & 1.57 & 4.27 \\
	0.5 &  4.096 & 3.0 &  -50.2  & 1.28 & 2.16 \\
	1.0 &  8.192 & 3.0 & -208.4  & 1.75 & 6.22 \\
	1.5 & 12.288 & 3.0 & -479.6  & 2.02 & 10.4 \\
	0.5 &  5.771 & 4.0 &  -98.6  & 1.71 & 5.74 \\
	1.0 & 11.542 & 4.0 & -410.9  & 2.24 & 15.1 \\
	1.5 & 17.312 & 4.0 & -945.3  & 2.56 & 24.8 \\
      \hline
    \end{tabular}
  \end{center}
  \caption{
    Summary of the numerical results from evolution of the truncated
    $1/r^2$ (left) and $\ln(r)/r^2$ (right) initial field profiles
    given by equations (\ref{eq:5D:IC}) and (\ref{eq:5D:IC:ln}) for
    various values of parameters. Black hole sizes and masses are
    approximate.
  }
  \label{tab:results}
\end{table*}

\begin{figure*}
  \begin{center}
    \begin{tabular}{cc}
      \epsfig{file=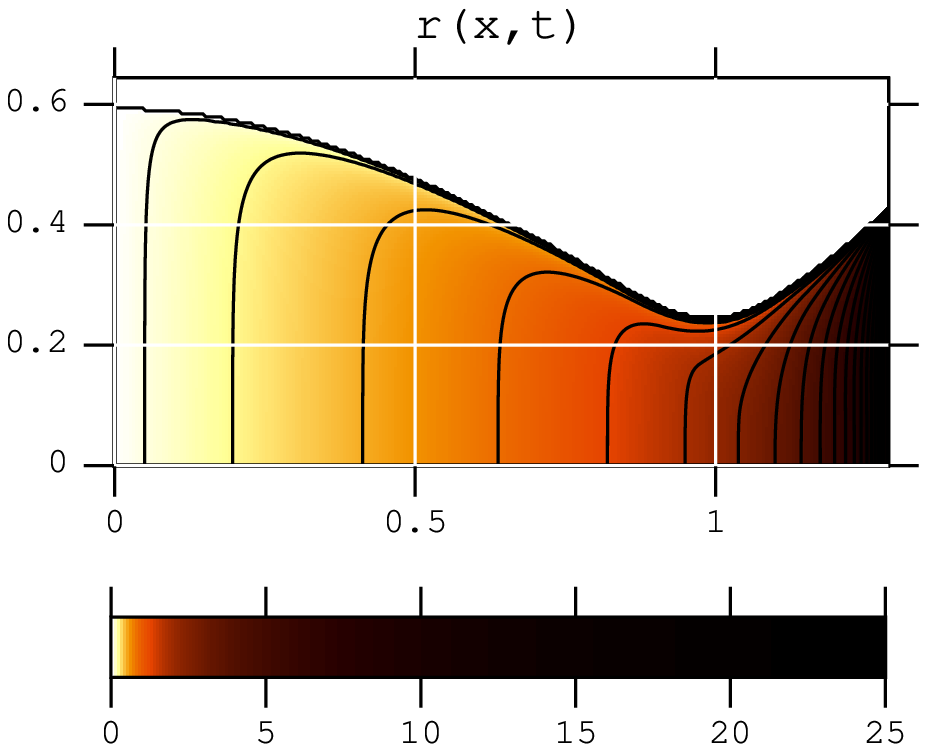, width=\columnwidth} &
      \epsfig{file=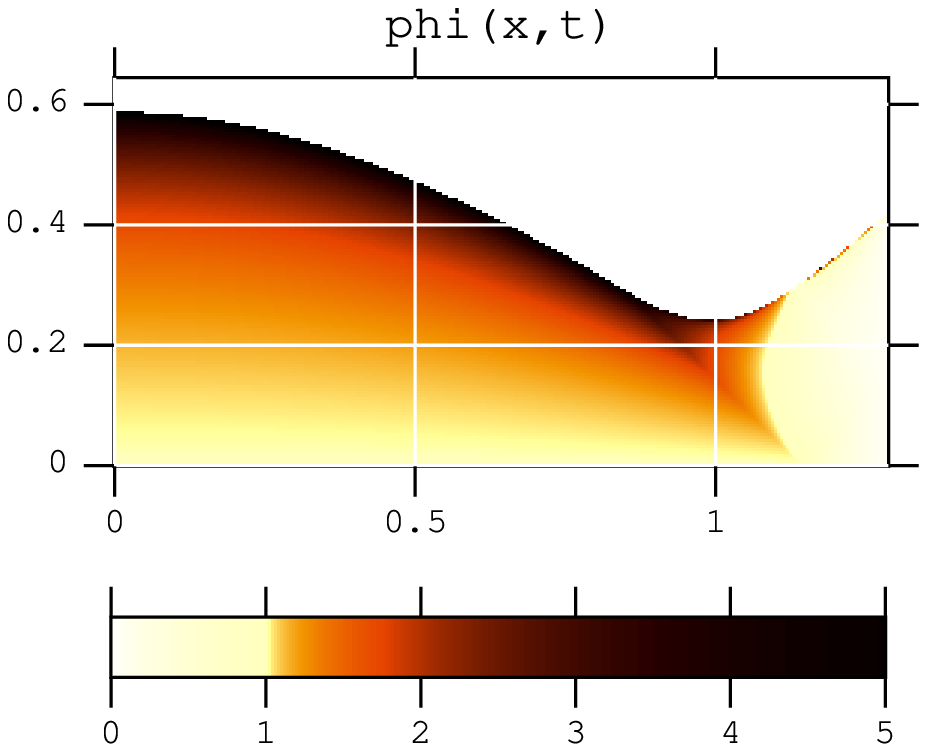, width=\columnwidth} \\
      \epsfig{file=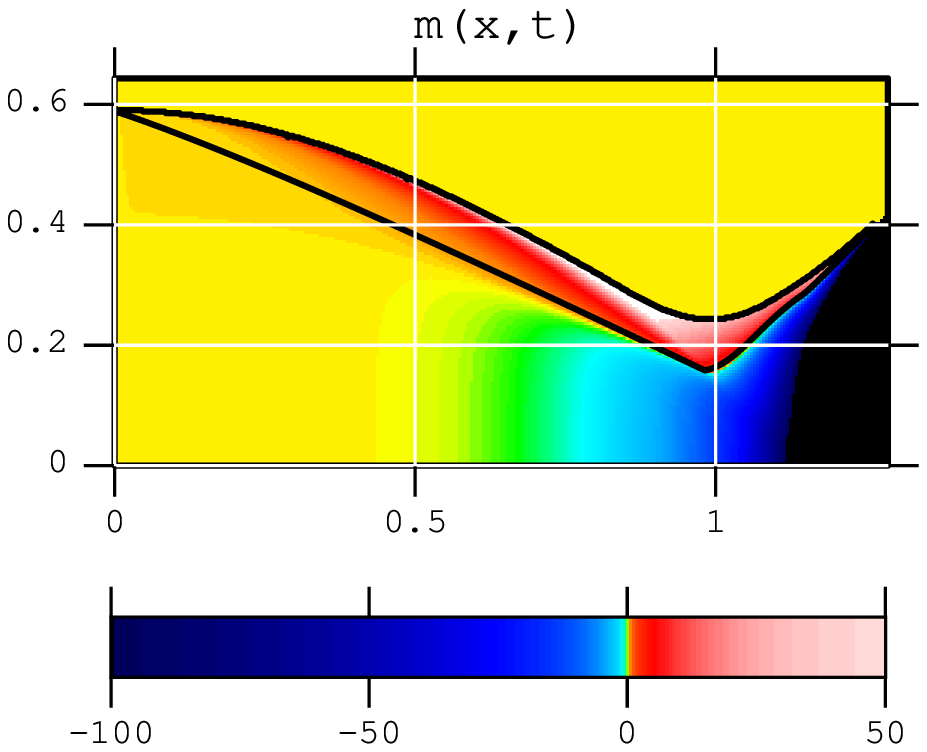, width=\columnwidth} &
      \epsfig{file=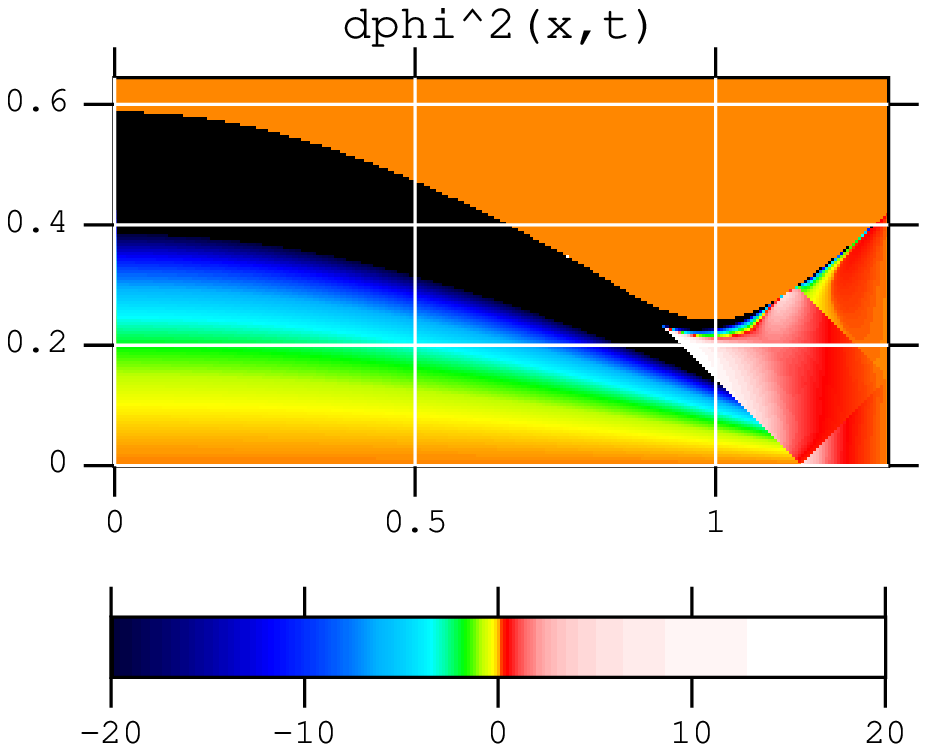, width=\columnwidth} \\
    \end{tabular}
  \end{center}
  \vspace{-5pt}
  \caption{
    Global structure of the spacetime resulting from the evolution of
    the truncated $1/r^2$ field profile (\ref{eq:5D:IC}) for the values
    of parameters $A=16$, $R_0=4$. Thick black lines on $m(x,t)$ plot
    show the locations of the apparent horizon and the singularity.
  }
  \label{fig:5D}
  \begin{center}
    \begin{tabular}{cc}
      \epsfig{file=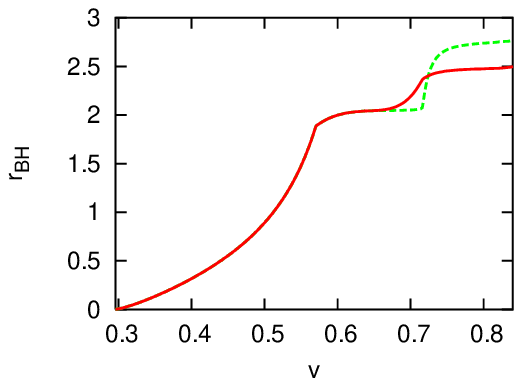, width=\columnwidth} &
      \epsfig{file=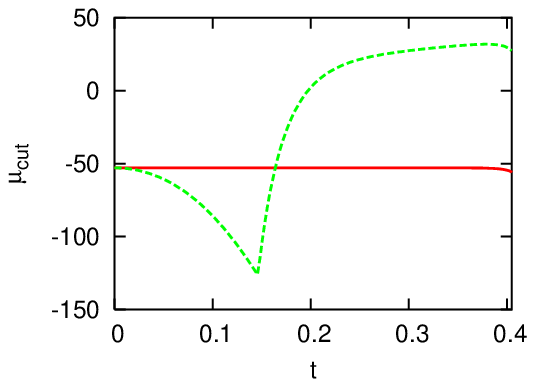, width=\columnwidth} \\
    \end{tabular}
  \end{center}
  \vspace{-10pt}
  \caption{
    The size of the apparent horizon (left) and the reduced mass $\mu$
    at the cut-off (right) for the same initial conditions as in
    Fig.~\ref{fig:5D}. Solid and dashed lines show the values
    corresponding to Dirichlet and Neumann boundary conditions at the
    cut-off.
  }
  \label{fig:horizon}
\end{figure*}

\begin{figure*}
  \begin{center}
    \begin{tabular}{cc}
      \epsfig{file=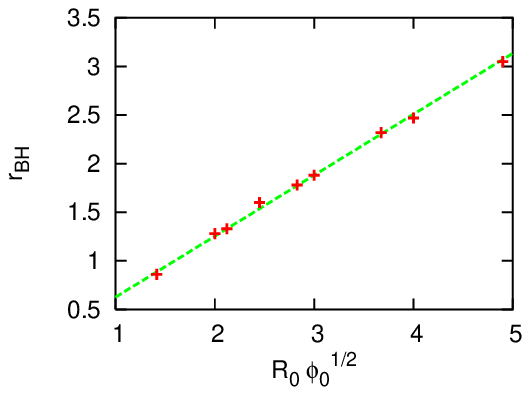, width=\columnwidth} &
      \epsfig{file=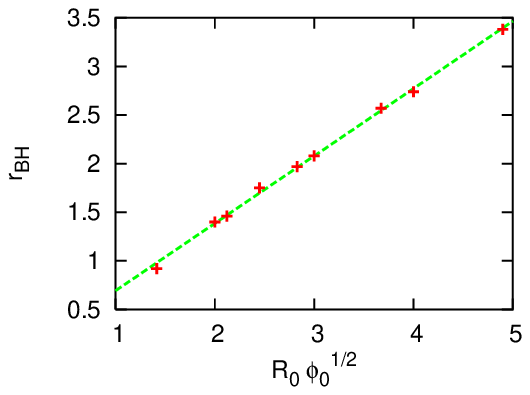, width=\columnwidth} \\
    \end{tabular}
  \end{center}
  \vspace{-10pt}
  \caption{
    The size of the black hole formed in the collapse of the truncated
    $1/r^2$ field profile (\ref{eq:5D:IC}) as a function of parameters,
    shown for Dirichlet (left) and Neumann (right) boundary conditions
    at the cut-off.
  }
  \label{fig:size}
\end{figure*}

Typical spacetime obtained in numerical simulations is shown in Fig.~\ref{fig:5D}.
It corresponds to evolution of initial field profile (\ref{eq:5D:IC})
with values of parameters $A=16$, $R_0=4$ ($\phi_0=1.0$) subject to
Dirichlet boundary conditions (\ref{eq:5D:BC:D}). The four panels show
density plots of metric coefficients $r(x,t)$ (top left) and $m(x,t)$
(bottom left), as well as the scalar field $\phi(x,t)$ (top right) and
its gradient $g^{\mu\nu}\phi_{,\mu}\phi_{,\nu}$ (bottom right).
Contours on $r(x,t)$ plot show isolines of constant radius. The upper
thick line on $m(x,t)$ plot is the singularity (where spacetime
evolution terminates), while the lower thick line is the apparent
horizon (found by condition $f=0$). Due to our coordinate choice
(\ref{eq:conf}), the plots in Fig.~\ref{fig:5D} essentially are
Carter-Penrose conformal diagrams, and show the global structure of the
spacetime. It is clear that the black hole has formed in the collapse.


To see if the black hole settles into an almost static configuration or
continues to evolve, we plot the size of the apparent horizon
$r_\text{BH}$ as a function of advanced time $v$ in the left panel of
Fig.~\ref{fig:horizon}. After the stage of growth corresponding to
collapse of the homogeneous part of the initial scalar field profile,
and sharp step-like increase in size when the outgoing portion of the
wavepacket reflects from infinity and falls in (these features in the
scalar field profile are most clearly seen in gradient plot in the
bottom right of Fig.~\ref{fig:5D}), the black hole settles to almost
constant size. The apparent horizon (shown in bottom left panel of
Fig.~\ref{fig:5D}) approaches null direction towards the end of the
evolution as well, which is consistent with the black hole settling
into a static configuration after the formation.

The formation of the black hole does not contradict the negative
initial reduced mass $\mu$ of the profile. For Dirichlet boundary
conditions (\ref{eq:5D:BC:D}), the reduced mass at the cut-off is
conserved, but the black hole {\em has scalar hair} which hides the
positive black hole mass and makes the total mass at the cut-off
negative. No-hair theorems \cite{Bekenstein:1995, Bronnikov:2001ah} do
not apply in this case as the cut-off is at a {\em finite} distance,
while the proof of the no-hair theorems involves exclusion of the
growing mode at infinity. For Neumann boundary conditions
(\ref{eq:5D:BC:N}), black holes do not have scalar hair, but the
reduced mass at the cut-off {\em is not conserved}, and grows positive
to accommodate the formation of the black hole, as shown in right panel
of Fig.~\ref{fig:horizon}.

In search of violation of cosmic censorship conjecture, we have
performed many high-resolution numerical simulations for different
initial profiles (\ref{eq:5D:IC}) and (\ref{eq:5D:IC:ln}), different
boundary conditions (\ref{eq:5D:BC:D}) and (\ref{eq:5D:BC:N}), and
various values of the parameters $\phi_0$ and $R_0$, but all in vain.
In no cases formation of naked singularity was seen. The global
structure of the spacetime remains similar to that of Fig.~\ref{fig:5D},
with black holes of various sizes formed in the collapse. Our numerical
results are summarized in Table~\ref{tab:results}, which shows sizes
and masses of the black holes produced. In addition, the spacetime mass
at the cut-off after black hole formation is shown for runs with
Neumann boundary conditions. The quoted values should be considered
approximate, as the number of useful grid points shrinks near the end
of the evolution and precision suffers somewhat (especially for smaller
black holes). It is worth noting that the black holes formed in the
runs with Neumann boundary conditions indeed have (almost) no hair, as
the masses evaluated at the horizon and at the cut-off are almost the
same.

As Fig.~\ref{fig:size} shows, the size of a black hole formed in the
collapse of the truncated $1/r^2$ field profile (\ref{eq:5D:IC}) scales
like $r_{\text{BH}} = \alpha R_0 \phi_0^{1/2} = \alpha A^{1/2}$ as a
function of profile parameters, with coefficient $\alpha \approx 0.627$
for Dirichlet and $\alpha \approx 0.693$ for Neumann boundary
conditions. This agrees well with the lower-bound estimate based on an
analysis of the collapse of the homogeneous part of the profile
\cite{Hubeny:2004cn}. Although the latter underestimates the value of
numerical coefficient (yielding $\alpha=0.58$), it is only by about 10-20\%.

\section{Conclusion}\label{sec:conlusion}

We have developed a double-null characteristic code implementing
$N$-dimensional spherically symmetric evolution of a minimally coupled
scalar field with potential in asymptotically AdS spacetimes, and used
it to study the possibility of cosmic censorship violation in string
theory. No instances of formation of naked singularity were seen in
high-resolution numerical simulations of the evolution of two families
of negative mass initial scalar field profiles for various values of
parameters and different boundary conditions at the cutoff.

Our results indicate that black holes form and reach steady-state
configuration in the collapse, despite the negative initial mass of
initial data. Either the spacetime mass becomes positive (for Neumann
boundary conditions, where mass $\mu$ is not conserved), or the black
hole covers itself with negative mass hair (for Dirichlet boundary
conditions, where no-hair theorems do not apply because of the finite
cutoff), as was pointed out in \cite{Hubeny:2004cn, Gutperle:2004jn}.

The possibility of naked singularity formation cannot be ruled out
completely by results of numerical simulations, as they only sample a
finite number of initial configurations, but in view of the above it
seems unlikely. Although no evolutions leading to big crunch were
observed from initial configurations considered in this paper, that
possibility should be explored further in the future work.

To summarize, the cosmic censorship conjecture was found to hold in all
the examples we have studied.

\bigskip
\section*{Acknowledgments}

I would like to thank Stephen Shenker, Veronika Hubeny, and Xiao Liu
for stimulating discussions. This work was supported by Kavli Institute
for Particle Astrophysics and Cosmology and Stanford Institute for
Theoretical Physics.
\vfill



\begin{thebibliography}{99}

\bibitem{Hertog:2003zs}
T.~Hertog, G.~T.~Horowitz and K.~Maeda,
{\it Generic cosmic censorship violation in anti de Sitter space},
Phys.\ Rev.\ Lett.\  {\bf 92}, 131101 (2004)
[{\tt gr-qc/0307102}].

\bibitem{Hertog:2003xg}
T.~Hertog, G.~T.~Horowitz and K.~Maeda,
{\it Negative energy in string theory and cosmic censorship violation},
Phys.\ Rev.\ D {\bf 69}, 105001 (2004)
[{\tt hep-th/0310054}].

\bibitem{Dafermos:2004ws}
M.~Dafermos,
{\it A note on naked singularities and the collapse of self-gravitating Higgs fields},
{\tt gr-qc/0403033}.

\bibitem{Hubeny:2004cn}
V.~E.~Hubeny, X.~Liu, M.~Rangamani and S.~Shenker,
{\it Comments on cosmic censorship in AdS/CFT},
{\tt hep-th/0403198}.

\bibitem{Gutperle:2004jn}
M.~Gutperle and P.~Kraus,
{\it Numerical study of cosmic censorship in string theory},
JHEP {\bf 0404}, 024 (2004)
[{\tt hep-th/0402109}].

\bibitem{Garfinkle:2004sx}
D.~Garfinkle,
{\it Numerical simulation of a possible counterexample to cosmic censorship},
Phys.\ Rev.\ D {\bf 69}, 124017 (2004)
[{\tt gr-qc/0403078}].

\bibitem{Garfinkle:2004pw}
D.~Garfinkle,
{\it Gravitational collapse in anti de Sitter space},
{\tt gr-qc/0408064}.

\bibitem{Hertog:2004gz}
T.~Hertog, G.~T.~Horowitz and K.~Maeda,
{\it Update on cosmic censorship violation in AdS},
{\tt gr-qc/0405050}.

\bibitem{Hertog:2004rz}
T.~Hertog and G.~T.~Horowitz,
{\it Towards a big crunch dual},
JHEP {\bf 0407}, 073 (2004)
[{\tt hep-th/0406134}].

\bibitem{Frolov:2003dk}
A.~V.~Frolov and U.~L.~Pen,
{\it The naked singularity in the global structure of critical collapse spacetimes},
Phys.\ Rev.\ D {\bf 68}, 124024 (2003)
[{\tt gr-qc/0307081}].

\bibitem{Bekenstein:1995}
J.~D.~Bekenstein,
{\it Novel `no scalar hair' theorem for black holes},
Phys.\ Rev.\ D {\bf 51}, 6608 (1995).

\bibitem{Bronnikov:2001ah}
K.~A.~Bronnikov and G.~N.~Shikin,
{\it Spherically symmetric scalar vacuum: No-go theorems, black holes and solitons},
Grav.\ Cosmol.\  {\bf 8}, 107 (2002)
[{\tt gr-qc/0109027}].

\end{thebibliography}

\end{document}